\documentclass[preprint2]{aastex}

\usepackage{epsf}

\slugcomment{{\em Accepted for publication the Astrophysical Journal Letters}} 

\begin{document}
\newcommand{\nhi}{\mbox{$N_{\rm HI}$}}
\newcommand{\hi}{H~{\sc i}}
\newcommand{\ahiss}{AH{\sc i}SS}
\newcommand{\himf}{H{\sc i}MF}
\newcommand{\mhi}{\mbox{$M_{\rm HI}$}}
\newcommand{\msol}{\mbox{${\rm M}_\odot$}}
\newcommand{\hmpc}{\mbox{$h_{100}^{-1}\, \rm Mpc$}}
\newcommand{\ihmpc}{\mbox{$h_{100}\, \rm Mpc^{-1}$}}
\newcommand{\kms}{\mbox{$\rm km\, s^{-1}$}}
\newcommand{\icmsq}{\mbox{$\rm cm^{-2}$}}
\newcommand{\rg}{\mbox{$r_{\rm g}$}}

\title{The Space Density of Primordial Gas Clouds near Galaxies and
Groups and their Relation to Galactic HVCs}

\author{Martin A. Zwaan  and Frank H. Briggs} 
\affil{Kapteyn Astronomical Institute, 
       Postbus 800, 
       9700 AV Groningen, 
       Netherlands\\
       zwaan@astro.rug.nl, fbriggs@astro.rug.nl}

\begin{abstract}
 The Arecibo \hi\ Strip Survey probed the halos of $\sim$300 cataloged
galaxies
and the environments of $\sim$14 groups with sensitivity to neutral
hydrogen masses $\geq 10^{7}~\msol$.  The survey detected no objects
with properties resembling the High Velocity Clouds (HVCs) associated
with the Milky Way or Local Group.  If the HVCs were typically
$\mhi=10^{7.5}~\msol$ objects distributed throughout groups and galaxy
halos at distances of $\sim$1~Mpc, the survey should have made $\sim$70
HVC detections in groups and $\sim$250 detections around galaxies.
The null detection implies that HVCs are deployed at typical distances
of $\leq$200~kpc from the galaxies or group barycenters.
 If the clouds are in virial equilibrium, their average dark matter
fraction must be 98\% or higher.
 \end{abstract}


\section{Introduction} 
 After an extensive review of the High Velocity Cloud (HVC) literature,
\citet{wak97} concluded that no single origin can
account for the properties of the HVC population of neutral hydrogen
clouds that surround the Milky Way Galaxy.  Instead, several mechanisms,
including infalling extragalactic clouds, cloud circulation within the
Galactic halo driven by a galactic fountain, and a warped outer arm
extension to the Galaxy must be invoked.  The Magellanic Stream and
associated HVC complexes require an explanation due to tidal
interactions within the Local Group (LG) \citep[cf.][]{put98}. 

A defining property for the HVCs has been the lack of associated stellar
emission.  This also means that spectroscopic parallax methods cannot be
used to measure distances to the clouds, and the lack of known
distances, in turn, hinders the calculation of the clouds' physical
properties.  Recently, absorption lines \citep[cf.][]{woe99} and Balmer
recombination line emission driven by reprocessing of the ionizing
radiation originating in the Galactic star forming regions
\citep{bla98,bla99} have been used to specify distances to a few of the
clouds, indicating a range of distances from within the Galactic halo to
greater than 50~kpc. 

Recent interest has returned to the idea that HVCs could be primordial
objects raining on the Galaxy, as either remnants from the formation of
the LG or as representatives from an intergalactic population of dark
matter dominated mini-halos in which hydrogen has collected and remained
stable on cosmological time scales (\citealt[BSTHB]{bli99};
\citealt[BB]{bra99}).  The requirement of gravitational stability
without star formation places lower limits on the distances of the
clouds from the Sun -- a sort of independent distance indicator that can
be applied to each cloud individually, depending on its \hi\ flux,
angular extent and velocity profile width.  Typical derived distances
are of order 1~Mpc, implying that this segment of the HVC population (1)
inhabits the LG rather than the Galactic halo, (2) has typical \hi\ mass
per cloud greater than ${\sim}10^7~\msol$, and (3) increases the LG \hi\
mass budget by contributing ${\sim}4{\times}10^{10}~\msol$ in the case
of the BSTHB sample.  For the BB sample, which is restricted to 65
confirmed ``compact HVCs'', the integral \hi\ content adds
${\sim}10^{9}~\msol$ to the LG. 

Further impetus to search for a primordial population of low mass
objects comes from simulations of galaxy and group formation
\citep[cf.][]{kly99}, which predict of order 10 times more
${\sim}10^8~\msol$ mini-halos in the LG than can be counted in
the dwarf galaxy population.  The association of HVCs with this missing
population, as well as arguments based on the kinematics of the cloud
population (BSTHB and BB), makes an appealing picture for the
extragalactic/LG explanation. 

Similar concerns motivated the extragalactic 21cm line survey by
\citet{wei91}, whose study of several representative environments
(clusters and voids) found only gas-rich galaxies containing stars. 
Several extragalactic \hi\ surveys of substantially larger volumes to
more sensitive \hi\ mass limits have also found no objects with HVC
properties
(i.e. \hi\ detections without associated starlight)
\citep{zwa97,spi99,kil99}. 

Clearly, if the HVC phenomenon is a common feature of galaxy formation
and evolution, then extragalactic surveys of the halos and group
environments of nearby galaxies should show evidence for this
population.  We take two approaches to the problem of placing the local
HVC population in an extragalactic context.  The first (in sections~3
and 4) is to compute the \hi\ mass function for the local group, both for
optically selected group members and for group members plus HVC
populations as modeled by BSTHB and BB.  The second, separate approach
(section~\ref{limits.sec}) is to calculate the probability that the
narrow strip that the Arecibo\footnote{The Arecibo Observatory is part
of the National Astronomy and Ionosphere Center, which is operated by
Cornell University under a cooperative agreement with the National
Science Foundation} \hi\ Strip Survey (\ahiss) \citep{sor94,zwa97} makes
through the halos of ${\sim}$200 galaxy halos and ${\sim}$14 groups
would detect members of HVC populations in those systems.

\section{The Local Group \hi\ Mass Function} \label{himf.sec}
 The \hi\ mass function (\himf) is used in extragalactic astronomy to
quantify the space density of gas-rich galaxies and possible
intergalactic clouds as a function of \hi\ mass.  For the field galaxy
population, the \himf\ has been determined accurately for
$\mhi>10^{7.5}~h_{65}^{-2}~\msol$ and can be fit satisfactorily with a
faint end slope $\alpha\approx -1.2$ \citep{bri93,zwa97,kil99}.  At
lower $\mhi<10^{7.5}~h_{65}^{-2}~\msol$, where there is considerable
uncertainty due to the small number of detections, \citet{sch98} have
found evidence for a steep upturn in the tail of the \himf.  Although
this steep tail has a tantalizing similarity to the signature of massive
HVCs, it appears that at least one of the two \hi\ signals responsible
for the rise comes from a normal galaxy, and the other is too close to a
bright star to exclude faint optical emission \citep{spi99}. 

 We construct the \himf\ for the LG from \hi\
measurements of all known LG members as compiled recently by
\citet{mat98}.  Within the LG volume, a total of 22 galaxies are
known in which \hi\ has been detected.  The statistics are therefore
poor, and consequently the \himf\ is noisy. 
 The LG \himf\ is in one sense the best measured \himf, with data over
six orders of magnitude, compared to three or four orders of magnitude
for determination for the field galaxy population.  On the other hand,
the LG \himf\ may suffer from severe selection effects due to
obscuration by dust in the disk of the Milky Way galaxy.  In order to
estimate how many galaxies might have escaped detection so far,
\citet{mat98} plotted the cumulative number of galaxies as a function of
$1-\sin|b|$, for the total sample of LG galaxies.  If LG galaxies were
distributed equally over the sky, the resulting histogram should be a
straight line.  We applied the same method to the sample of galaxies
with \hi\ detections and found that 4 to 7 galaxies with \hi\ masses
$>10^4~\msol$ are likely to be missing at low Galactic latitude.  The
missing galaxies would be predominantly the ones with low optical
surface brightness, but since there is no clear correlation between
surface brightness and \hi\ mass, it is not possible to make a more
refined correction to the \himf\ than just adding one galaxy
to each half decade mass bin below $\mhi=10^8~\msol$.  For larger
galaxies with $\mhi>10^8~\msol$, we assume that the census of the LG
galaxy population is complete \citep[cf.][]{hen98}.

\begin{figure}
\resizebox{\hsize}{13cm}{\includegraphics{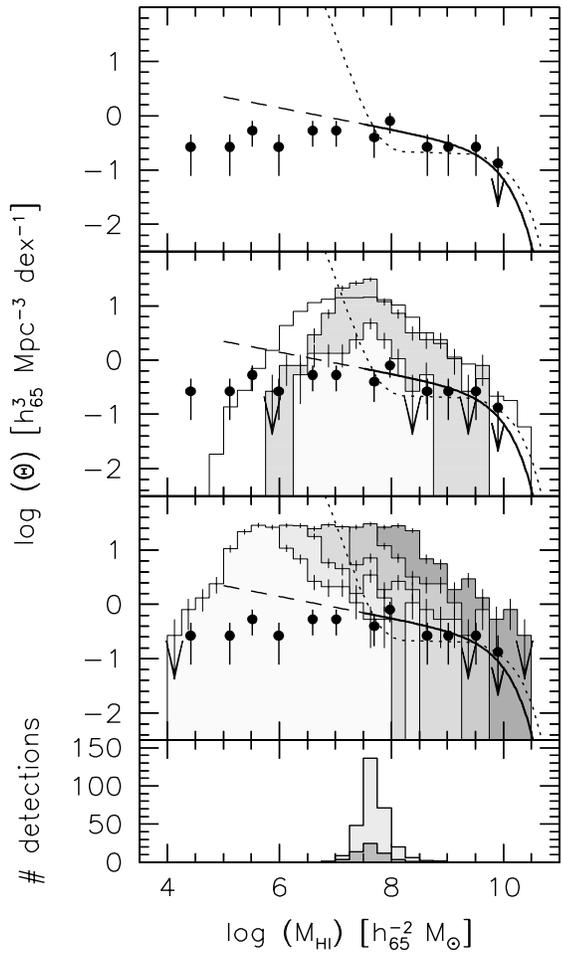}}
\vspace{-0.1cm}
\caption{
\footnotesize {\em Top panel:\/} \hi\ mass function of the Local Group
(LG).  The
points show the space density of LG members containing \hi, after
correcting for incompleteness.  The solid line shows the field \himf\
from Zwaan et al.  (1997), scaled vertically so as to fit the points. 
The dotted line is a \himf\ with a steep upturn at the low mass end,
recently proposed by Schneider et al.  (1998)
 {\em Second panel:\/} \hi\ mass functions for extragalactic HVCs.  The
dark histogram shows the space density of BSTHB HVCs if they are put at
the critical radii for gravitational stability.  The unshaded histogram
shows the effect of a baryon to total mass ratio $f$ varying from cloud
to cloud.  The lighter shaded histogram corresponds to BB HVCs, all at
the same distance of 1~Mpc.  Lines are as in top panel.
  {\em Third panel:\/} \hi\ mass function for BSTHB HVCs for different
values $f$ (from right to left: $f$=0.2, 0.1, 0.05, 0.025, 0.0125).  The
HVC \himf\ is consistent with the field \himf\ if $f\leq 0.02$ and the
median distance $\leq 200$~kpc.
  {\em Bottom panel:\/} Distribution of the expected \hi\ detections in
the \ahiss\ for the BSTHB population.  The light grey histogram
corresponds to clouds surrounding galaxies, the dark grey histogram
represents clouds in galaxy groups.  See section~4 for explanation.
\label{fig1.fig}}
\end{figure}

 The result is presented in the top panel of Fig.~\ref{fig1.fig},
where the points represent the LG \himf\ and the line is the field
\himf\ derived by \citet{zwa97} scaled vertically so as to fit the
points in the region around $\mhi=10^9~\msol$ where the curve has been
measured accurately.  
 This scaling is justified given the fact that the \himf\ for optically
selected and \hi\ selected galaxies is identical \citep[cf.][]{bri93,zwa97}.
The scaling accounts for the overdensity of the
LG, which amounts to a factor of 25, assuming that the LG volume is 15
Mpc$^3$.  
Also shown is a \himf\ with a steep upturn proposed by \citet{sch98}.  The
large divergence between the extrapolated curves from \citet{zwa97} and
\citet{sch98} illustrates the uncertainty in \himf\ below 
$\mhi=10^{7.5}~\msol$.
The LG \himf\ for optically selected galaxies is
remarkably flat, with the faint-end slope of a Schechter function fit of
$\alpha \approx -1.0$. 

 Studies of the \hi\ content of galaxies in different environments
(including voids [\citealt{szo96}], clusters [\citealt{mcm93}] and 
groups [\citealt{kra99}]) have
shown that the shape of the \himf\ for $\mhi > 10^8~\msol$ is
independent of cosmic density.   The
fact that we find here that the \himf\ of optically selected LG members
is flat down to \hi\ masses of a few $\times~10^4~\msol$ does not, however,
insure that the field \himf\ is also flat to these low masses. 
Although the crossing time of the LG is approximately equal to a Hubble
time, there are clear indications of interactions \citep[and references
therein]{mat98}.  The \hi\ distributions in the lowest luminosity LG
dwarfs are often highly asymmetric, perhaps indicative of tidal
distortions.  It is quite possible that low mass systems are destroyed
or merged, which could cause the LG \himf\ to be flatter than that of
the field.

\section{\hi\ Mass Functions for Extragalactic HVCs}\label{hvchimfs.sec}
 We now derive H{\sc i}MFs for the population of extragalactic
HVCs as proposed by BSTHB and BB.  
For the compact, isolated HVCs identified by BB,
we estimate \hi\ masses by using their measurements of
the integrated fluxes and the assumption that all clouds are at the same
1~Mpc distance, suggested by BB.  Typical \hi\ masses are then a few
times $10^7~\msol$ and the sizes are approximately 15 kpc.  The
resulting \himf\ is indicated by the light shaded histogram
in the second panel of Fig.~\ref{fig1.fig}.  For the BSTHB sample we use
the compilation of HVCs by \citet[WW]{wak91}, which forms the main
source for the BSTHB analysis.  For each cloud the distance \rg\ at
which the cloud is gravitationally stable is calculated.  Adopting
BHTHB's value of $f=0.1$ for the ratio of baryonic mass to total mass
leads to typical distances of self-gravitating HVCs of $\sim 1$~Mpc. 

The dark shaded histogram represents the \himf\ for the BSTHB clouds,
assuming that all clouds are at their distance \rg.  The resulting \himf\
shows a clear peak at approximately $3\times 10^7~\msol$,
equal to the typical \hi\ mass estimated by BSTHB.  The largest
uncertainty in the determination of \mhi\ is $f$, which might vary from
cloud to cloud.  We tested the effect of a varying $f$ on the \himf\ by
applying to each individual cloud a random value of $f$ in the range
$0.03$ to $0.3$. The resulting \himf\ (shown by the unshaded histogram)
is not significantly different from the \himf\ based on a fixed $f$ in
the region of interest $>10^7\msol$.

 The space density of the BB HVC population is comparable to the \himf\
for optically selected LG galaxies, but it is obvious that the BSTHB
HVCs outnumber normal galaxies by a factor 5 to 10 in the range
$10^{7.5} < \mhi < 10^9~\msol$.  This implies that if the BSTHB cloud
population is typical for galaxy groups, \hi\ surveys in groups should
have encountered 5 to 10 dark \hi\ clouds for every detected galaxy. 
This is clearly at variance with the observations. 

At what distances from the Milky Way must the HVCs be located so that
their \himf\ is not in conflict with the observed field \himf? Since the
virial distance \rg\ is directly proportional to $f$, we can test this
by varying $f$ from 0.2 to 0.025 and calculate the resulting \himf.  The
results are shown in the third panel of Fig.~\ref{fig1.fig}.  The space
density of HVCs in the LG can only be brought into agreement with the
observed field \himf\ if the median value of $f$ is lowered to
$\sim~0.02$, a value much lower than what is normally observed in
galaxies.  The median distance of such clouds must be smaller than $\sim
200$ kpc.

\section{Expected number of extragalactic HVC detections}
\label{limits.sec}
\begin{figure}[ht]   
\resizebox{\hsize}{!}{\includegraphics{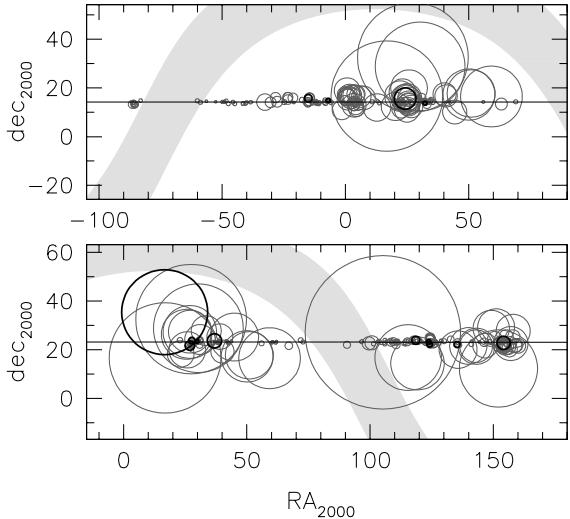}}
\vspace{-0.1cm} 
\caption{\footnotesize Illustration of the two surveys strips and the distribution of spheres
with radii of 1 Mpc around known galaxies (thin circles) and galaxy
groups (thick circles).  The grey areas indicate the Zone of Avoidance
where $|b|<10^\circ$.  The solid horizontal lines show the paths of the
Arecibo beam.  \label{fig2.fig}}
\end{figure}

 The Arecibo \hi\ Strip Survey (\ahiss), which is discussed in detail in
\citet{sor94} and \citet{zwa97}, puts limits on the space density of
primordial gas clouds in external galaxy groups and around galaxies. 
The survey was taken in drift-scan mode and consists of two strips at
constant declinations, together covering $20^{\rm h}$ of RA over a
redshift range from $0$ to $7500$ \kms.  The limiting column density was
$10^{18}~\icmsq$, which is lower than that of most of the HVCs in the WW
compilation and those presented in BB.  The sky coverage is small (15
deg$^2$ excluding the side lobes) but the survey strip
passes through the halo regions
of many groups and galaxies as shown in Fig.~\ref{fig2.fig}.  
The unique character of this survey makes
it therefore more suitable to assess the HVC problem than other surveys
of equal size.    From the
Lyon-Meudon Extragalactic Database (LEDA\footnote{We have made use of
the Lyon-Meudon Extragalactic Database (LEDA) supplied by the LEDA team
at the CRAL-Observatoire de Lyon (France). }) we selected all known
galaxies with projected distances $<1 h^{-1}_{65}$~Mpc from the two
\ahiss\ strips.  The circles in Fig.~\ref{fig2.fig}
indicate shells with 1~Mpc radii around the
galaxies.  Since the discussion in BSTHB is primarily focussed on galaxy
groups, we also selected all cataloged groups within 1~Mpc of the
strips.  Galaxy groups were drawn from \citet{wil97}, who used the the
Mark~III catalog, and \citet{gar93} who selected groups from the LEDA
galaxy sample.

We fill the volumes around the selected groups and galaxies with a
synthetic population of HVCs similar to the one proposed in BSTHB.  To
construct such an ensemble we make use of the compilation of HVCs by WW
as discussed in section~\ref{hvchimfs.sec}.  Although BSTHB put
particular emphasis on galaxy groups, we choose to consider clouds around
individual galaxies as well.  Hierarchical formation scenarios do not
distinguish between galaxies and groups in the relative number of
satellites \citep{kly99}.  Further motivation comes from the fact that
in the LG, subclustering of dwarfs is observed around the Milky Way and
M31 \citep{mat98}. 

The $3'$ beam of the Arecibo telescope subtends $d_{\rm beam}=0.87 D
~\rm kpc$ at a distance $D$ Mpc.  The typical sizes of the HVCs
discussed in BSTHB are 28 kpc and the lowest column density clouds could
therefore be detected out to distances of 32~Mpc, beyond which the
average HVC would no longer fill the beam.  HVCs with column densities
in excess of the limiting value of $10^{18}~\icmsq$ could be detected to
larger distances.  The limiting column density for clouds at large
distances where $d_{\rm cloud}<d_{\rm beam}$ is $N_{\rm
lim}=10^{18}\times(d_{\rm beam}/d_{\rm cloud})^2$.  For each group and
galaxy, a fraction of the volume of the surrounding sphere is scanned by
the Arecibo beam.  The number of clouds within that volume is
calculated, taking into account the column densities and sizes of the
individual clouds. 

Table~\ref{table1.tab} lists the number of clouds that would have been
detected in the Arecibo \hi\ strip survey if a population of
extragalactic HVCs existed with the BSTHB properties.  
We calculate the numbers
of clouds differently for groups and for galaxies.  Since BSTHB do
not specify the exact radial distribution of clouds, we tested three
different radial distribution functions to fill the
volumes with clouds:
 1) a spherical volume of radius $R$,
 2) a thin spherical shell of radius $R$, and
 3) a thick spherical shell with clouds distributed according to a
Gaussian about the radius $R$ with dispersion $\sigma=R/3$. 
 The latter distribution most closely resembles the derived
distribution of $r_{\rm g}$ given in BSTHB.  
The numbers in the table are based on $R=1~\rm Mpc$, the value 
preferred by both BSTHB and BB.
The group
halos are filled with 450 clouds, the number of HVCs identified by 
WW, excluding complexes A, C, M, the Outer Arm and the Magellanic Stream.
To calculate the expected number of clouds around galaxies, the
number of clouds associated with each galaxy is scaled in direct
proportion to the ratio of the galaxy luminosity compared to the
integral luminosity of the LG.  This leads to a median number of clouds
per galaxy of 40. 

Table~\ref{table1.tab} shows that the expected number of detections is
essentially independent of how the clouds are distributed around the
groups and galaxies.  For these samples we should detect approximately
250 clouds around galaxies and 70 around groups if the HVC population of
BSHTB is a common feature of nearby galaxies.  Restricting our
analysis to the compact clouds of BB, reduces these numbers
to 39 and 9.  For a uniformly filled spherical distribution of
BSTHB clouds, the distribution of \hi\ masses of the expected detections
is shown in the lowest panel in Fig.~\ref{fig1.fig}.  This figure
illustrates that our analysis is sensitive to typical \hi\ clouds
(compare second panel of Fig.~\ref{fig1.fig}), and not only to the
most massive ones. 

The robustness of the result is demonstrated in Table~\ref{table1.tab},
where the numbers in parentheses indicate the expected number of
detections if the detection threshold is increased from $5\sigma$ to
$7\sigma$.  The average decrease is 25\%.  A 50\% decrease occurs if the
detection threshold is set at $10\sigma$.  This extremely conservative
threshold would still predict more than 100 detections.

\section{Conclusions}\label{discussion.sec}
 The hypothesis that HVCs are primordial gas clouds with typical \hi\
masses of a few $\times 10^7~\msol$ at distances of $\sim 1$~Mpc from
the Galaxy is not in agreement with observations of nearby galaxies and
groups. 
 Blind \hi\ surveys of the extragalactic sky would have detected
these clouds if they exist around all galaxies or galaxy groups in
numbers equal to those suggested for the Local Group.  These results are
highly significant: the Arecibo \hi\ strip survey would have detected
approximately 250 clouds around individual galaxies and 70 in galaxy
groups.  

\acknowledgments We are grateful to B.  Wakker for discussions and for
providing the list of HVC parameters.  J.  Bland-Hawthorn, L.  Blitz,
R.  Braun, J.  van Gorkom, and H.  van Woerden are thanked for useful
comments.

\clearpage
\begin{deluxetable}{lrrrrr}
\tablecolumns{6}
\tablewidth{0pc}
\tablecaption{Expected extragalactic HVCs detections}
\tablehead{
 \colhead{} & \multicolumn{3}{c}{BSTHB} & \colhead{} & \colhead{} \\
 \cline{2-4} \\
 \colhead{} & \colhead{uniform} & \colhead{shell} & \colhead{Gaussian} &
 \colhead{BB} & \colhead{selected} \\
}
\startdata
groups      & 70(52)    & 72(54)     & 70(52)   & 9(8)     & 14   \\
galaxies    & 248(167)  & 256(173)   & 260(177)  & 39(28)   & 347  \\
 \tablecomments{The number of expected HVC detections are calculated
assuming a $5\sigma$ detection threshold.  The numbers in parentheses
indicate the expected number if the threshold is raised to $7\sigma$.   
The last column shows the number of groups and galaxies that are
included in the analysis.  For the Gaussian distributions a selection
radius of 2 Mpc has been used for inclusion in the sample.
 \label{table1.tab} }
 \enddata
 \end{deluxetable}

\end{document}